\documentclass[10pt,conference]{IEEEtran}

\usepackage{booktabs} 
\usepackage{xcolor,colortbl}
\definecolor{Gray}{gray}{0.9}
\definecolor{Green}{RGB}{89,163,76}
\usepackage{fontawesome}
\usepackage{subfigure}
\usepackage{url}
\usepackage{todonotes}
\usepackage{balance}
\usepackage{framed}
\usepackage{textcomp}
\usepackage{amssymb}
\usepackage{microtype}
\usepackage{xfrac}

\usepackage[numbers,sort&compress]{natbib}

\newcommand{\highlight}[1]{\begin{framed}%
  \noindent\emph{#1}
\end{framed}}

\begin{document}
\title{EMaaS: Energy Measurements as a Service for Mobile Applications}

\author{\IEEEauthorblockN{Luis Cruz}
\IEEEauthorblockA{University of Porto\\
INESC-TEC\\
Porto, Portugal\\
\url{luiscruz@fe.up.pt}}
\and
\IEEEauthorblockN{Rui Abreu}
\IEEEauthorblockA{Instituto Superior T\'ecnico, University of Lisbon\\
INESC-ID\\
Lisbon, Portugal\\
\url{rui@computer.org}}
}

\maketitle

\begin{abstract}

Measuring energy consumption is a challenging task faced by developers when
building mobile apps. This paper presents EMaaS: a system that provides
reliable energy measurements for mobile applications, without requiring a complex
setup. It combines estimations from an energy model with --- typically more
reliable, but also expensive --- hardware-based measurements. On a per scenario
basis, it decides whether the energy model is able to provide a reliable
estimation of energy consumption. Otherwise, hardware-based measurements are
provided. In addition, the system is accessible to the community of mobile
software practitioners/researchers in the form of a Software as a Service. With
this service, we aim at solving current problems in the field of energy
efficiency in mobile software engineering: the complexity of hardware-based
power monitor tools, the reliability of energy models, and the continuous need
of data to build energy models. \end{abstract}


\section{Introduction}
\label{sec:intro}

One of the main problems mobile developers face when building apps is the
need to run tests under a number of different settings (e.g., mobile device
models, operative system
version)~\citep{moran2017automated,muccini2012software}. For instance, the
appearance of the user interface may change in devices with different screen
sizes and resolutions. This problem is also evident when testing energy
consumption.

\begin{figure}
  \centering
\includegraphics[width=0.6\linewidth]{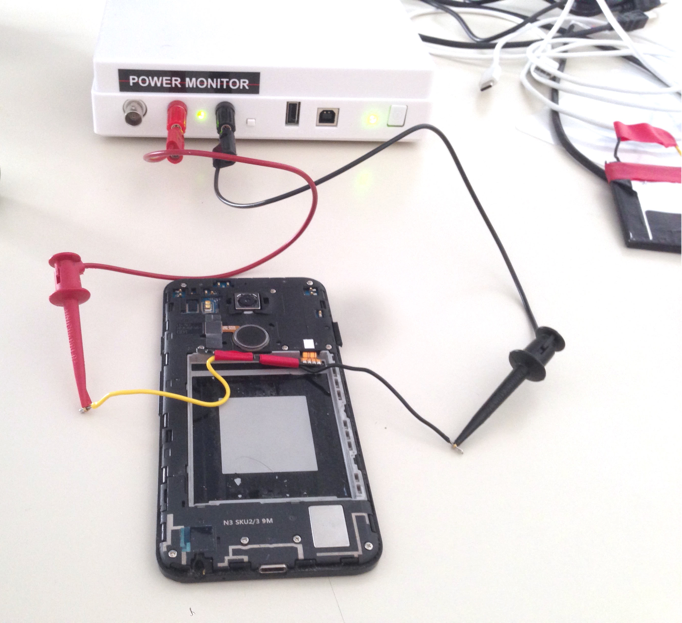}
\caption{\emph{Monsoon} power monitor connected to the Nexus 5X smartphone.}
\label{fig:monsoon}
\vspace{-1.5em}
\end{figure}

Moreover, measuring energy consumption with power tools is a cumbersome task.
Developers need to create a setup that often requires disassembling the
device and acquire specialized power tools, as shown in
Fig.~\ref{fig:monsoon}. Such setup is expensive, time-consuming and requires
skills out of the domain of most practitioners. Nevertheless, this setup ends up
being unused most of the time, since developers do not need to run energy
measurements continuously during development.

To overcome these issues, tools have been proposed to measure
energy consumption using software-based
estimators~\citep{hao2013estimating,di2017petra,chowdhury2018greenscaler,zhang2010accurate}. These tools are able to provide accurate estimations within the
context in which they were trained~\citep{di2017software}. However, when they
are used against new settings --- for instance, different devices, operative
system (OS) versions, or API methods --- one cannot be sure about the reliability
of those measurements. Thus, there is a need to continuously collect energy
measurement data in order to have accurate energy models.

Although energy models are very handy, practitioners cannot be sure of whether
a model is trained for a given use case. A developer should not rely on a model
that was trained for different scenarios. However, this information is not
always accessible to developers. This is critical when developers need to
make design decisions based on the energy efficiency of their code.

There is a tradeoff between using hardware and software-based energy
estimators. None of these approaches is ready to be adopted by the community of
mobile developers on a global scale. In this paper, we address the
aforementioned issues by proposing a hybrid system that provides the
best of these two approaches. The system provides energy measurements as a
service and is able to switch between software and hardware-based measurements,
depending on the app under test and its running environment.

In this paper, we propose EMaaS, a peer-to-peer cloud-based system that delivers
energy measurements as a service for mobile applications. In particular, the
system addresses the following issues:

\begin{enumerate}

  \item Power monitor tools are complex and impractical in a
real mobile development scenario.

  \item Reliable energy models need to be
continuously updated with new data, collected using hardware-based power monitors.

  \item The context in which a given energy model can provide reliable
  measurements is not always clear.

\end{enumerate}

\section{The Vision}

We envisage a crowd-sourced system to deliver lightweight energy tests as a
service. It combines energy models with power monitors to provide the most
accurate energy measurements without requiring a cumbersome setup of power
tools. Moreover, the system allows developers to measure their apps in mobile
devices from different manufacturers.

Developers only need to provide an executable package with (1) the application
build (e.g. an \emph{APK}\footnote{APK is the package file format used by the
Android OS for distribution and installation of mobile apps and middleware.} in
the case of the Android OS) and (2) the instrumentation build with the test
cases to be measured (e.g., the instrumentation \emph{APK} for \emph{Espresso}
test cases on Android).

\begin{figure}
\includegraphics[width=\linewidth]{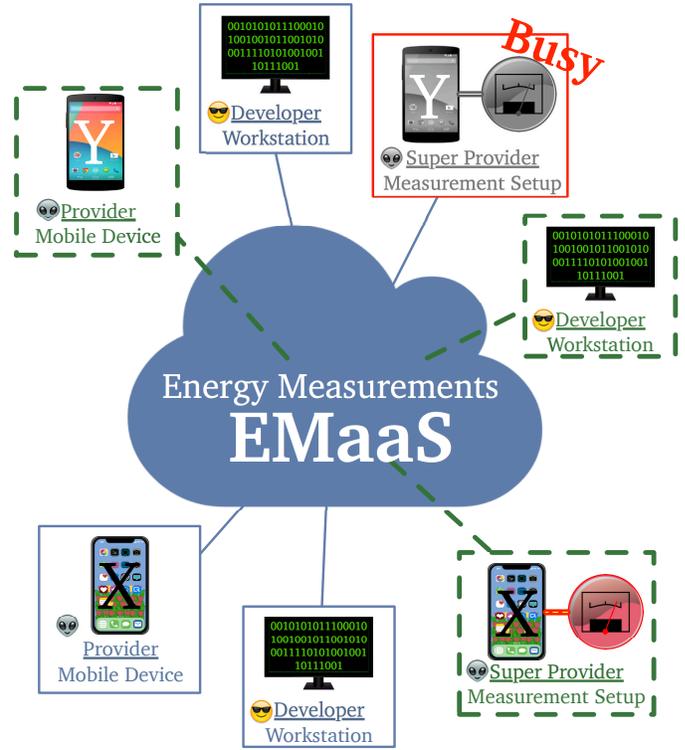}
\caption{High-level vision of the system for energy measurements as a service.}
\vspace{-1.5em}
\label{fig:arch}
\end{figure}

Fig.~\ref{fig:arch} outlines the high-level architecture of the system.
It features three types of users:

\begin{itemize}

 \item\textbf{Developers} use the system to collect energy consumption
measurements for their apps.

 \item\textbf{Providers} allow other users to use their mobile devices to run
energy measurements. In this case, energy consumption is estimated using an
energy model.

 \item\textbf{Super-Providers} are special \textbf{providers} that provide
energy measurements using a hardware-based power monitor.

\end{itemize}

The same user can act as a developer, provider, or super-provider,
simultaneously. When super-providers are not available to run a given
measurement, providers take their place using energy models. 

As an example, Fig.~\ref{fig:arch} highlights three connected peers, marked
with green thick lines: one developer, one provider, and one super-provider. The
same developer asks the system to run energy tests in two different device
models, X and Y. The system assigns a super-provider for device model X but no
super-provider was available for device Y. Although a super-provider for Y
exists in the system, it is busy dealing with another measurement task. Thus,
since the system had a reliable energy estimator for device model Y, it
assigned the task to a provider, as depicted in the illustration. If the system
did not have a suitable energy model for the given scenario, the task would
wait until a super-provider for model Y would become available.

In this setting, the developer does not own any mobile device or power monitor
tool. They are being provided by other users in the system. With this approach,
it delivers 1) a better use of resources, 2) access to a bigger set of
devices and tools, and 3) a simple/affordable setup to measure energy
consumption.

Under the hood, the system integrates two modules to bring the best of both
hardware and software-based power monitors: the \textbf{Energy Model} and the
\textbf{Reliability Consultant}.

The \textbf{Energy Model} is based on approaches from previous work, which uses
data collected from hardware-based power monitors to train models of power
consumption. The main limitation of current state-of-the-art solutions is the
lack of data to train energy models~\citep{chowdhury2018greenscaler}. We
mitigate this limitation by continuously updating models using data collected
from hardware-based power monitors (i.e., super-providers). Thus, in this
solution, providers always estimate energy consumption using up-to-date energy
models.

Another limitation is the fact that these solutions rely on in-vitro data to
train their models. Extrapolating these estimators to different devices can
compromise measurements. On contrary, this solution can virtually scope any
execution environment, given that it is available from a
\textbf{super-provider}. In addition, since the power model is continuously
improving every time developers run their energy tests, it is able to adapt to
new paradigms, devices, or OS versions when they come to market. Going back to
the example given by Fig.~\ref{fig:arch}, if a reliable energy model for the
given test cases was not available for device Y, the system would collect data
from super-providers over time. Eventually, the system would have enough data
to learn a new energy model that could provide reliable estimations for the new
scenario.

Although energy models can provide very accurate estimations, one needs to
assess whether they are ready to be used in a given context. I.e., a given
energy model might not be ready yet to measure the energy consumption of apps that
use specific libraries. The system needs to assess whether, for a given
developer request, it is acceptable to return an energy estimation. If not, the
system has to alert the developer or wait for a super-provider to be available.
We propose the module \textbf{Reliability Consultant} to solve this problem.

In parallel with the \textbf{Energy Model}, the \textbf{Reliability Consultant}
module will inspect the reliability of the estimator. Super-providers will run
simultaneously hardware and software-based measurements. The reliability of
energy models is assessed using results from hardware as ground-truth. We then
construct reliability as a metric that is negatively correlated to the power
error ($\epsilon$), given by the following equation:

\begin{equation}
  \epsilon = \frac{E_{measured} - E_{estimated}}{\Delta t}
\end{equation}

\noindent where $E_{measured}$ and $E_{estimated}$ are the energy consumption
measured by the power monitor and the estimator, respectively, and $\Delta t$
is the duration of execution of the energy test. For high reliability,
$\epsilon$ should be close to zero.

In addition, we use static analysis to collect data regarding the app and its
execution environment --- for instance, frequency of library/API calls,
complexity metrics, framework, API level, OS version, etc. This data will be
used to train a regression model that estimates the power error of the
software-based estimator.

As a side contribution we propose to answer the following research questions:

\highlight{RQ1: Can static analysis be used to assess the reliability of energy
models for different running environments? }


\highlight{RQ2: What is the improvement of using hybrid energy measurements
over software-based measurements? }

\highlight{RQ3: What is the proportion of software vs. hardware measurements? }

\begin{figure}
  \centering
\includegraphics[width=0.9\linewidth]{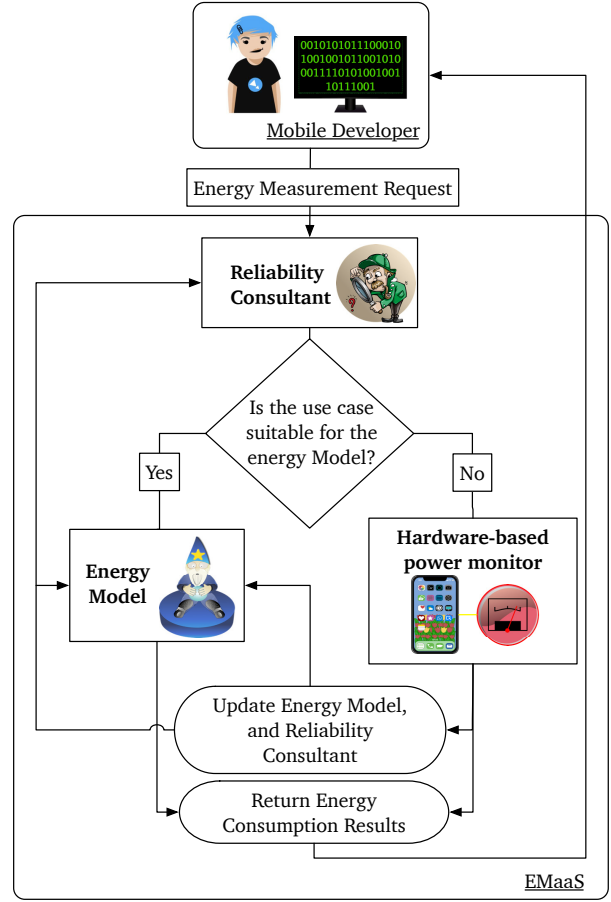}

\caption{How a normal energy measurement request is processed, starting from the
perspective of a developer.}

\label{fig:decision}
\vspace{-1.5em}
\end{figure}

The combination of the modules \textbf{Reliability Consultant}, \textbf{Energy
Model}, and \textbf{Hardware-based Power Monitor} is depicted in
Fig.~\ref{fig:decision}. Upon a measurement request from the developer, the
system asks the \textbf{Reliability Consultant} whether the measurement needs
to use a \textbf{Hardware-based Power Monitor}. If not, the estimation provided
by the \textbf{Energy Model} is returned to the developer. On contrary, if the
\textbf{Reliability Consultant} does not consider the \textbf{Energy Model}
reliable for the given mobile app, an hardware-based measurement is provided,
and the \textbf{Energy Model} and \textbf{Reliability Consultant} are updated
with new data.

\section{Why is it New?}

Very interesting tools have been proposed by researchers to measure energy
consumption. However, given the ever-changing nature of the mobile application
world, keeping them up-to-date and ready to use for developers can be
challenging. This system is able to be continuously updated and to adapt its
energy model to new settings.

Moreover, this will help researchers having more valuable contributions in the
field of energy efficiency of mobile applications. Recent contributions are
using energy models to estimate
energy consumption~\citep{li2014empirical,li2015optimizing,sahin2014code,chowdhury2018exploratory}.
However, the validity of these measurements is not easily assessed. In
addition, most contributions have their experiments limited to a small set of
environments, providing a relevant threat to the validity.

This work bridges the gap between industry and academia in terms of energy
measurements for mobile applications. To the best of our knowledge, this is the
first time reliable energy measurements can be deployed to the industry without
requiring an expensive setup. Furthermore, no prior work has leveraged a tool
to assess the compatibility between an energy model and its running environment.

\section{Risks}

The main risks that can affect this system are related to \emph{Security} and
\emph{Privacy}.

\paragraph{Security}

The system has to account for malware mobile software that can affect the
devices of providers.

\paragraph{Privacy}

Developers do not want to disclosure their apps before deployment. Mechanisms
need to be deployed to prevent mobile applications from being collected by
malicious providers or super-providers. Furthermore, the mobile devices
available in the system need to be protected from mal-intended developers. The
data stored in the devices should not be accessed by the system or the mobile
application under test.

\paragraph{Measurement best practices}

Typically, energy measurements require rigorous approaches to mitigate
potential bias on measurements. For instance, the app under analysis should be
running in isolation on the phone while no other apps are running in
parallel~\citep{cruz2017performance}.

To address these issues, energy tests have to be executed in a closed
environment, instantiated in the mobile device. In some cases, the device may
have to be restored. Thus, providers may not be able to use their own personal
devices. For safety reasons, the system may be limited to devices exclusively
acquired for app development.

\section{Next Steps}

In this paper, we present the idea of using peer-to-peer cloud computing to
deliver reliable energy measurements as a service. We are interested in
improving the way researchers and developers are profiling the energy
consumption of their mobile software. By making it available as service, we aim
to help developers make informed design decisions regarding the energy
efficiency of their code.

In the early stages, EMaaS will require a considerable software development
effort. Thus, we divide the implementation into three steps. In the first step,
we will be interested in delivering a system that yields hardware-based
measurements. This will serve as a base system to add more sophisticated energy
measurement techniques in the following steps.

The second step will consist of including the \textbf{Energy Model}. In particular,
we will be looking at how to improve existing models for energy estimation. One
big advantage of our approach is the access to real measurement data collected
from a wide variety of devices and OSs.

Finally, in the third step, we will be adding the \textbf{Reliability
Consultant}. We will use measurements collected from both energy models and
power monitors to train a regression model of the reliability of energy
estimators in a particular execution scenario.

For an initial proof-of-concept, the system will support Android apps
instrumented with \emph{Espresso} test cases, since it has been the most
suitable UI test framework for energy
measurements~\citep{cruz2018measuring,cruz2019attention}. In order to initially
avoid privacy and security risks, the system will be exclusively available to
invited users.

\section{Acknowledgments}

Luis Cruz is sponsored by the ERDF -- European Regional Development Fund through
the Operational Programme for Competitiveness and Internationalisation -
COMPETE 2020 Programme and by National Funds through the Portuguese funding
agency, FCT -- Fundação para a Ciência e a Tecnologia, within project
POCI-01-0145-FEDER-016718, and by the FCT scholarship grant
number PD/BD/52237/2013.

Rui Abreu is sponsored by the ERDF through the COMPETE 2020 Program and by
National Funds through the Portuguese funding agency FCT with reference
UID/CEC/50021/2019, and within the project FaultLockerRef
(PTDC/CCI-COM/29300/2017).

\balance
\bibliographystyle{IEEEtranN}
\bibliography{bibliography}

\end{document}